# Attenuation Coefficient Estimation for PET/MRI With Bayesian Deep Learning pseudo-CT and Maximum Likelihood Estimation of Activity and Attenuation

Andrew P. Leynes, Sangtae Ahn, Kristen A. Wangerin, Sandeep S. Kaushik, Florian Wiesinger, Thomas A. Hope, and Peder E. Z. Larson

*Abstract*— A major remaining challenge for magnetic resonance-based attenuation correction methods (MRAC) is their susceptibility to sources of MRI artifacts (e.g. implants, motion) and uncertainties due to the limitations of MRI contrast (e.g. accurate bone delineation and density, and separation of air/bone). We propose using a Bayesian deep convolutional neural network that, in addition to generating an initial pseudo-CT from MR data, also produces uncertainty estimates of the pseudo-CT to quantify the limitations of the MR data. These outputs are combined with MLAA reconstruction that uses the PET emission data to improve the attenuation maps. With the proposed approach (UpCT-MLAA), we demonstrate accurate estimation of PET uptake in pelvic lesions and show recovery of metal implants. In patients without implants, UpCT-MLAA had acceptable but slightly higher RMSE than Zero-echo-time and Dixon Deep pseudo-CT when compared to CTAC. In patients with metal implants, MLAA recovered the metal implant; however, anatomy outside the implant region was obscured by noise and crosstalk artifacts. Attenuation coefficients from the pseudo-CT from Dixon MRI were accurate in normal anatomy; however, the metal implant region was estimated to have attenuation coefficients of air. UpCT-MLAA estimated attenuation coefficients of metal implants alongside accurate anatomic depiction outside of implant regions.

*Index Terms*— Bayesian deep learning, deep learning, MLAA, MRAC, synthetic CT

## I. Introduction

The quantitative accuracy of simultaneous positron emission tomography and magnetic resonance imaging (PET/MRI) depends on accurate attenuation correction. Simultaneous imaging with positron emission tomography and computed tomography (PET/CT) is the current clinical gold standard for PET attenuation correction since the CT images can be used for attenuation correction of 511keV photons with piecewise-linear models [1]. Magnetic resonance imaging (MRI) measures spin density rather than electron density and thus cannot directly be used for PET attenuation correction.

A comprehensive review of attenuation correction methods for PET/MRI can be found at [2]. Briefly, current methods for attenuation correction in PET/MRI can be grouped into the following categories: atlas-based, segmentation-based, and machine learning-based. Atlas-based methods utilize a CT atlas that is generated and registered to the acquired MRI [3]–[6]. Segmentation-based methods use special sequences such as ultrashort echo-time (UTE) [7]–[11] or zero echo-time (ZTE) [12]–[16] to estimate bone density and Dixon sequences [17]–[19] to estimate soft tissue densities. Machine learning-based methods, including deep learning methods, use sophisticated machine learning models to learn mappings from MRI to pseudo-CT images [20]–[26] or PET transmission images [27]. There have also been methods that estimate attenuation coefficient maps from the PET emission data [28], [29] or directly correct PET emission data [30]–[32] using deep learning.

For PET alone, an alternative method for attenuation correction is "joint estimation", also known as maximum likelihood estimation of activity and attenuation (MLAA) [33], [34]. Rather than relying on an attenuation map that was measured or estimated with another scan or modality, the PET activity image ($\lambda$-map) and PET attenuation coefficient map ($\mu$-map) are estimated jointly from the PET emission data only. However, MLAA suffers from numerous artifacts and high noise [35].

In PET/MRI, recent methods developed to overcome the limitations of MLAA include using MR-based priors [36], [37], constraining the region of joint estimation [38], or using deep learning to denoise the resulting $\lambda$-map and/or $\mu$-map from MLAA [39]–[42]. Mehranian and Zaidi's [36] approach of using priors improved MLAA results however this was not demonstrated on metal implants. Ahn et al and Fuin et al's methods [37], [38] that also use priors were able to recover metal implants in the PET image reconstruction, but the $\mu$-maps

This paper was submitted for review on August 16, 2021. This project was supported in part by NIH/NCI R01CA212148, NIH/NIAMS R01AR074492, the UCSF Graduate Research Mentorship Fellowship award, and GE Healthcare. The Titan X Pascal used was donated by the NVIDIA Corporation. (*Corresponding author: Andrew P. Leynes*)

A. P. Leynes, T. A. Hope, and P. E. Z. Larson are with the Department of Radiology and Biomedical Imaging, University of California San Francisco, San Francisco, CA 94158 USA (e-mail: Andrew.Leynes@ucsf.edu).

A. P. Leynes and P. E. Z. Larson are also with the UC Berkeley – UC San Francisco Joint Graduate Program in Bioengineering, Berkeley and San Francisco, CA USA

S. Ahn is with GE Research, Niskayuna, NY USA

K. A. Wangerin is with GE Healthcare, Waukesha, WI USA

S. S. Kaushik and F. Wiesinger are with GE Healthcare, Munich, Germany

S. S. Kaushik is also with the Department of Computer Science, Technical University of Munich, Munich, Germany

T. A. Hope is also with the Department of Radiology, San Francisco VA Medical Center, San Francisco, CA, USA

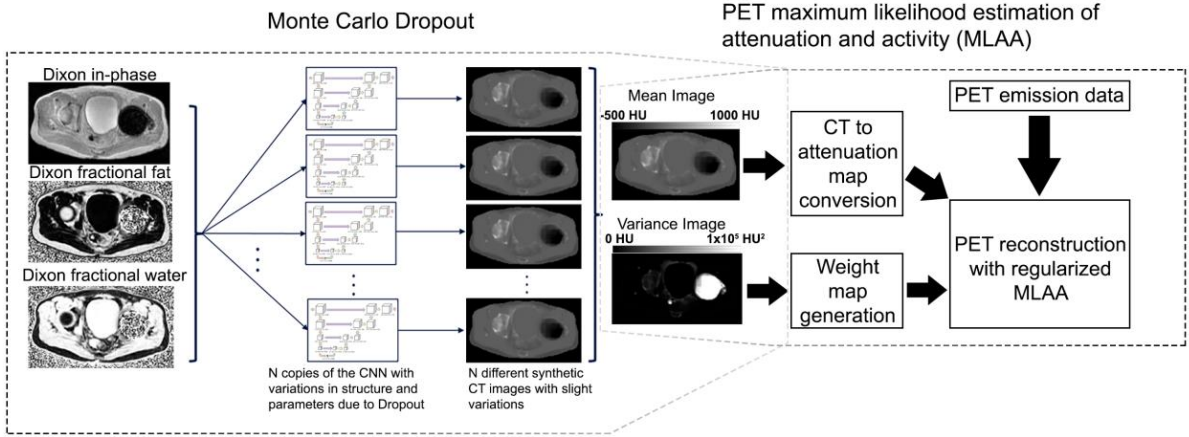

Fig 1. Schematic flow of UpCT-MLAA. Monte Carlo Dropout is first performed with the BCNN, then the outputs are provided as inputs to PET reconstruction with regularized MLAA.

were missing bones and other anatomical features. Furthermore, their methods require a manual or semi-automated segmentation step to delineate the regions where to apply the correct priors (such as the metal implant region). The approaches by Hwang et al [39]–[41] and Choi et al [42] that utilize supervised deep learning resulted in anatomically correct and accurate $\mu$-maps; however, the method was not demonstrated in the presence of metal implants.

Utilizing supervised deep learning is considered a very promising method for accurate and precise PET/MRI attenuation correction. However, the main limitation of a supervised deep learning method is the finite data set that needs to have a diverse set of well-matched inputs and outputs.

In PET/MRI, the presence of metal implants complicates training because there are resulting metal artifacts in both CT and MRI. Furthermore, the artifacts appears differently: a metal implant produces a star-like streaking pattern with high Hounsfield unit values in the CT image [43] and a signal void in the MRI image [37]. This makes registration between MRI and CT images difficult and the artifacts lead to intrinsic errors in the training dataset.

In addition, there will arguably always be edge cases and rare features that cannot be captured with enough representation in a training data set. Images of humans can have rare features not easily obtained (e.g., missing organs due to surgery, a new or uncommon implant). Under these conditions, a standard supervised deep learning approach may produce incorrect predictions and the user (or any downstream algorithm) will be unaware of the errors.

A recent study by Ladefoged et al [44] demonstrated the importance of a high-quality data set in deep learning-based brain PET/MRI attenuation correction. A large, diverse set of at least 50 training examples were required to achieve robustness and they highlighted that the remaining errors and limitations in deep learning-based MR attenuation correction were due to "abnormal bone structures, surgical deformation, and metal implants."

In this work, we propose the use of supervised Bayesian deep learning to estimate *predictive* uncertainty to detect rare or previously unseen image structures and estimate intrinsic errors that traditional supervised deep learning approaches cannot.

Bayesian deep learning provides tools to address the limitations of a finite training dataset: the estimation of *epistemic* and *predictive* uncertainty [45]. A general introduction to uncertainties in machine learning can be found at [46].

Epistemic uncertainty is the uncertainty on learned model parameters that arises due to incomplete knowledge or, in the case of supervised machine learning, the lack of training data. Epistemic uncertainty is manifested as a diverse set of different model parameters that fit the training data.

The *epistemic* uncertainty of the model can then be used to produce *predictive* uncertainty that capture if there are any features or structures that deviate from the training dataset on a test image. This allows for the detection of rare or previously unseen image structures without explicitly training to identify these structures.

Typical supervised deep learning approaches do not capture the *epistemic* nor *predictive* uncertainty because only one set of model parameters are learned and only a single prediction is produced (e.g., a single pseudo-CT image).

In this work for PET/MRI attenuation correction, the predictive uncertainty is used to automatically weight the balance between the deep learning $\mu$-map prediction from MRI and the $\mu$-map estimates from the PET emission data from MLAA. When the model is expected to have good performance on a region in a test image, then MLAA has minimal contribution. However, when the model is expected to have poor performance on regions in a test image, then MLAA has a stronger contribution to the attenuation coefficient estimates of those regions.

Specifically, we extend the framework of Ahn et al's MLAA regularized with MR-based priors [37] and generate MR-based priors with a Bayesian convolutional neural network (BCNN) [47] that additionally provides a predictive uncertainty map to automatically modulate the strength of the MLAA priors. We demonstrate a proof-of-concept methodology that produces anatomically correct, accurate, and precise $\mu$-maps with high SNR that can recover metal implants for PET/MRI attenuation correction in the pelvis.

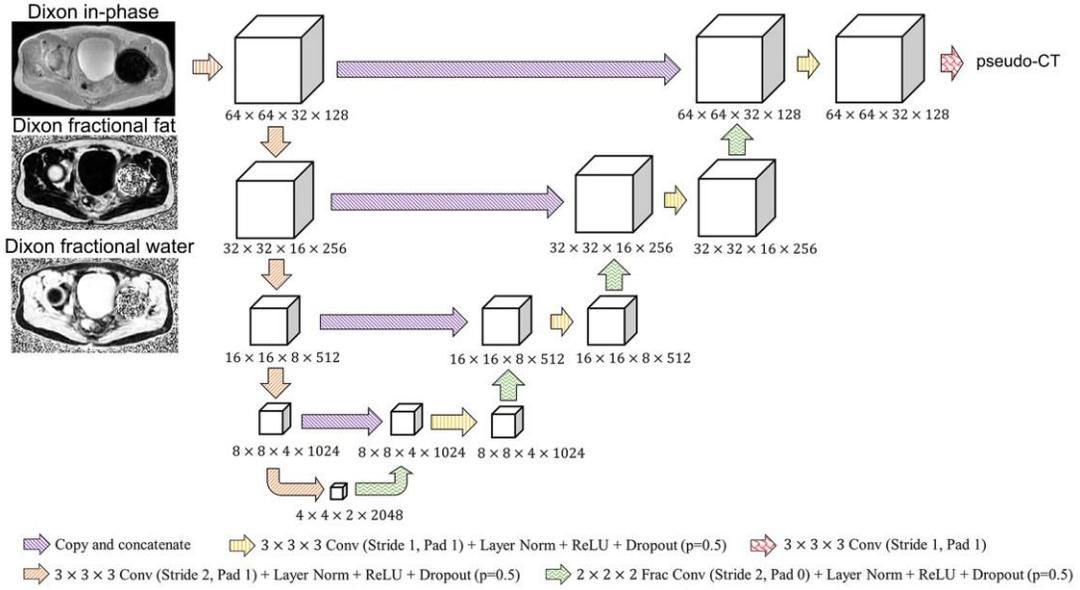

Fig 2. Deep convolutional neural network architecture used in this work.

## II. MATERIALS AND METHODS

UpCT-MLAA is composed of two major elements: initial pseudo-CT characterization with Bayesian deep learning through Monte Carlo Dropout [47] and PET reconstruction with regularized MLAA [37]. The algorithm is depicted in Fig. 1 and each component is described in detail below.

### A. Bayesian Deep Learning

The architecture of the BCNN is shown in Fig. 2. It was based on the U-net-like network in [21] with the following modifications: (1) Dropout [47], [48] was included after every convolution, (2) the patch size was increased to $64 \times 64 \times 32$ voxels, and (3) each layer's number of channels was increased by 4 times to compensate for the reduction of information capacity due to the Dropout. The PyTorch software package [49] (v0.4.1, http//pytorch.org) was used.

Inputs to the model were volume patches of the following dimensions and size: 64 pixels × 64 pixels × 32 pixels × 3 channels. Each channel was a volume patch of the bias-corrected and fat-tissue normalized Dixon in-phase image, Dixon fractional fat image, and Dixon fractional water image, respectively, at the same spatial locations [50]. The output was a corresponding pseudo-CT image with size 64 pixels × 64 pixels × 32 pixels × 1 channel. ZTE MRI was not used as inputs to this model since it has been demonstrated that accurate HU estimates can be achieved with only the Dixon MR pulse sequence [22], [50].

#### 1) Model Training

Model training was performed similarly to our previous work [21], [50]. The loss function was a combination of an L1-loss, gradient difference loss (GDL), and Laplacian difference loss (LDL):

$$Loss = |y - \hat{y}| + \lambda_{GDL}\left(|\nabla_x y - \nabla_x \hat{y}|^2 + |\nabla_y y - \nabla_y \hat{y}|^2 + |\nabla_z y - \nabla_z \hat{y}|^2\right) + \lambda_{LDL}(|\Delta y - \Delta \hat{y}|^2) \quad (1)$$

where $\nabla$ is the gradient operator, $\Delta$ is the Laplacian operator, $y$ is the ground-truth CT image patch, and $\hat{y}$ is the output pseudo-CT image patch with $\lambda_{GDL} = 0.01$ and $\lambda_{LDL} = 0.01$. The Adam optimizer [51] (learning rate = $1 \times 10^{-5}$, $\beta_1 = 0.9$, $\beta_2 = 0.999$, $\epsilon = 1 \times 10^{-8}$) was used to train the neural network. An L2 regularization ($\lambda = 1 \times 10^{-5}$) on the weights of the network was used. He initialization [52] was used and a mini-batch of 4 volumetric patches was used for training on two NVIDIA GTX Titan X Pascal (NVIDIA Corporation, Santa Clara, CA, USA) graphics processing units. The models were trained for approximately 68 hours to achieve 100,000 iterations.

### B. Pseudo-CT prior and weight map

Generation of the pseudo-CT estimate and variance image was performed through Monte Carlo Dropout [47] with the BCNN described above. The Monte Carlo Dropout inference is outlined in Fig. 1. A total of 243 Monte Carlo samples were performed to generate a pseudo-CT estimate and a variance map:

$$pCT = \frac{1}{N}\sum_{i=1}^{N} f_i(x) \quad (2)$$

$$\sigma^2 = \frac{1}{N}\sum_{i=1}^{N}(f_i(x) - pCT)^2 \quad (3)$$

where $f_i$ is a sample of the BCNN with Dropout, $x$ is the input Dixon MRI, and N is the number of Monte Carlo samples. Inference took approximately 40 minutes per patient on 8 NVIDIA K80 graphics processing units. We include a detailed description of the sources of uncertainties and variations in the Supplementary Material.

The pseudo-CT estimate was converted to a $\mu$-map with a bilinear model [1] and the variance map was converted to a

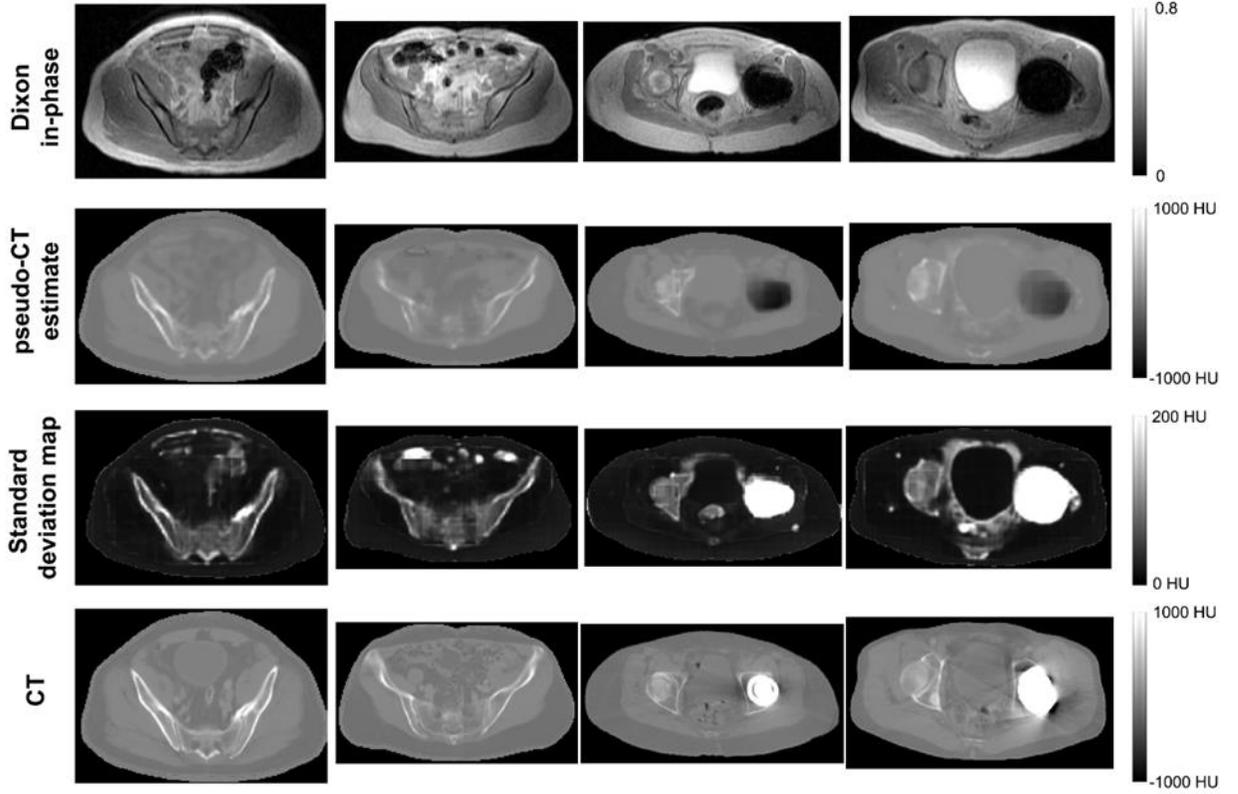

Fig 3. Representative intermediate image outputs of the BCNN with Monte Carlo Dropout compared to the reference CT images for patients without metal implants (columns 1 and 2) and patients with metal implants (columns 3 and 4). The voxel-wise standard deviation map is shown instead of variance for better visual depiction. Regions with high standard deviation correspond to bone, bowel air, skin boundary, implants, blood vessels, and regions with likely modeling error (e.g. around the bladder in the standard deviation map in the rightmost column.)

weight map with a range of 0.0 to 1.0 with the following empirical transformation:

$$\mathbf{w}(\vec{r}) = \frac{1}{1 + \exp\left(0.1\left(\left(\frac{\sigma^2(\vec{r})}{1000}\right) - 25\right)\right)} \quad (4)$$

where $\sigma^2(\vec{r})$ is the variance at voxel position $\vec{r}$. The sigmoidal transformation was calibrated by inspecting the resulting variance maps. It was designed such that the transition band of the sigmoid covers the range of variances in the body and finally saturates at uncertainty values of bowel air and metal artifact regions. With the constants chosen, the transition band of the sigmoid corresponds to variances of 0 to ~100,000 HU² (standard deviations of 0 to ~300 HU). The weight map was then linearly scaled to have a range of $1 \times 10^3$ to $5 \times 10^6$, called $\beta_{MR}$. The low $\beta_{MR}$ values correspond to regions with high uncertainty and thus the estimation for these regions would be dominated by the emission data. Additional information about the empirical transformation is provided in the Supplementary Material.

The weight map was additionally processed to set weights outside the body (e.g. air voxels) to 0.0 so that these were not included in MLAA reconstruction. A body mask was generated by thresholding (> -400 HU) the pseudo-CT estimate. The initial body mask was morphologically eroded by a 1-voxel radius sphere. Holes in the body were then filled in with the *imfill* function (Image Processing Toolbox, MATLAB 2014b) at each axial slice. The body masks were then further refined by removing arms as in our previous work [14].

*C. Uncertainty estimation and pseudo-CT prior for robust Maximum Likelihood estimation of Activity and Attenuation (UpCT-MLAA)*

UpCT-MLAA is a combination of the outputs of the BCNN and regularized MLAA. The process is depicted in Fig. 1. MRI and CT images of patients without metal implants were used to train the BCNN.

We explicitly trained the network only on patients without metal implants to force the BCNN to extrapolate on the voxel regions containing metal implant (i.e. "out-of-distribution" features) to maximize the uncertainty in these regions.

Thus, a high variance (>=~$1 \times 10^5$ HU²) emerged in implant regions compared to a low variance in normal anatomy (0 to ~$2.5 \times 10^4$ HU²) with the uncertainty estimation as can be seen in Fig. 1. The $\mu$-map estimate and the weight map were then provided to the regularized MLAA [37] to perform PET reconstruction (5 iterations with 28 subsets, each iteration consists of 1 TOF-OSEM iteration and 5 ordered subsets transmission (OSTR) iterations, $\beta_{MR}$ as described above, $\beta_{smooth}=2 \times 10^4$). Specifically, the MR-based regularization term in MLAA is:

$$R_{MR}(\mu) = \sum_i \frac{\beta_{MR_i}}{2}(\mu_i - \mu_i^{MR})^2 \quad (5)$$

where $i$ indexes over each voxel in the volume. $\mu^{MR}$ is determined from the mean pseudo-CT image and $\beta_{MR}$ is determined from the variance image through the weight map transformation. The formulation in eq. 5 is slightly different than in Section 2.3.2 of [37] but has the same effect.

III. PATIENT STUDIES

The study was approved by the local Institutional Review Board (IRB). Patients who were imaged with PSMA-11 signed a written informed consent form while the IRB waived the requirement for informed consent for FDG and DOTATATE studies.

Patients with pelvic lesions were scanned using an integrated 3 Tesla time-of-flight PET/MRI system [53] (SIGNA PET/MR, GE Healthcare, Chicago, IL, USA). The patient population consisted of 29 patients (Age = $58.7 \pm 13.9$ years old, 16 males, 13 females): 10 patients without implants were used for model training, 16 patients without implants were used for evaluation with a CT reference, and three patients with implants were used for evaluation in the presence of metal artifacts.

### A. PET/MRI Acquisition.

The PET acquisition on the evaluation set was performed with different radiotracers: $^{18}$F-FDG (11 patients), $^{68}$Ga-PSMA-11 (7 patients), $^{68}$Ga-DOTATATE (1 patient). The PET scan had 600 mm transaxial field-of-view (FOV) and 25 cm axial FOV, with time-of-flight timing resolution of approximately 400 ps. The imaging protocol included a six bed-position whole-body PET/MRI and a dedicated pelvic PET/MRI acquisition. The PET data were acquired for 15-20 min during the dedicated pelvis acquisition, during which clinical MRI sequences and the following MRAC sequences were acquired: Dixon (FOV = $500 \times 500 \times 312$ mm, resolution = $1.95 \times 1.95$ mm, slice thickness = 5.2 mm, slice spacing = 2.6 mm, scan time = 18 s) and ZTE MR (cubical FOV = $340 \times 340 \times 340$ mm, isotropic resolution = $2 \times 2 \times 2$ mm, 1.36 ms readout duration, FA = 0.6°, 4 μs hard RF pulse, scan time = 123 s).

### B. CT Imaging.

Helical CT images of the patients were acquired separately on different machines (GE Discovery STE, GE Discovery ST, Siemens Biograph 16, Siemens Biograph 6, Philips Gemini TF ToF 16, Philips Gemini TF ToF 64, Siemens SOMATOM Definition AS) and were co-registered to the MR images using the method outlined below. Multiple CT protocols were used with varying parameter settings (110-130 kVp, 30-494 mA, rotation time = 0.5 s, pitch = 0.6-1.375, 11.5-55 mm/rotation, axial FOV = 500-700 mm, slice thickness = 3-5 mm, matrix size = 512×512).

Pre-processing consisted of filling in bowel air with soft-tissue HU values and copying arms from the Dixon-derived pseudo-CT due to the differences in bowel air distribution and the CT scan being acquired with arms up, respectively [14].

MRI and CT image pairs were co-registered using the ANTS [54] registration package and the SyN diffeomorphic deformation model with combined mutual information and cross-correlation metrics [14], [21], [50].

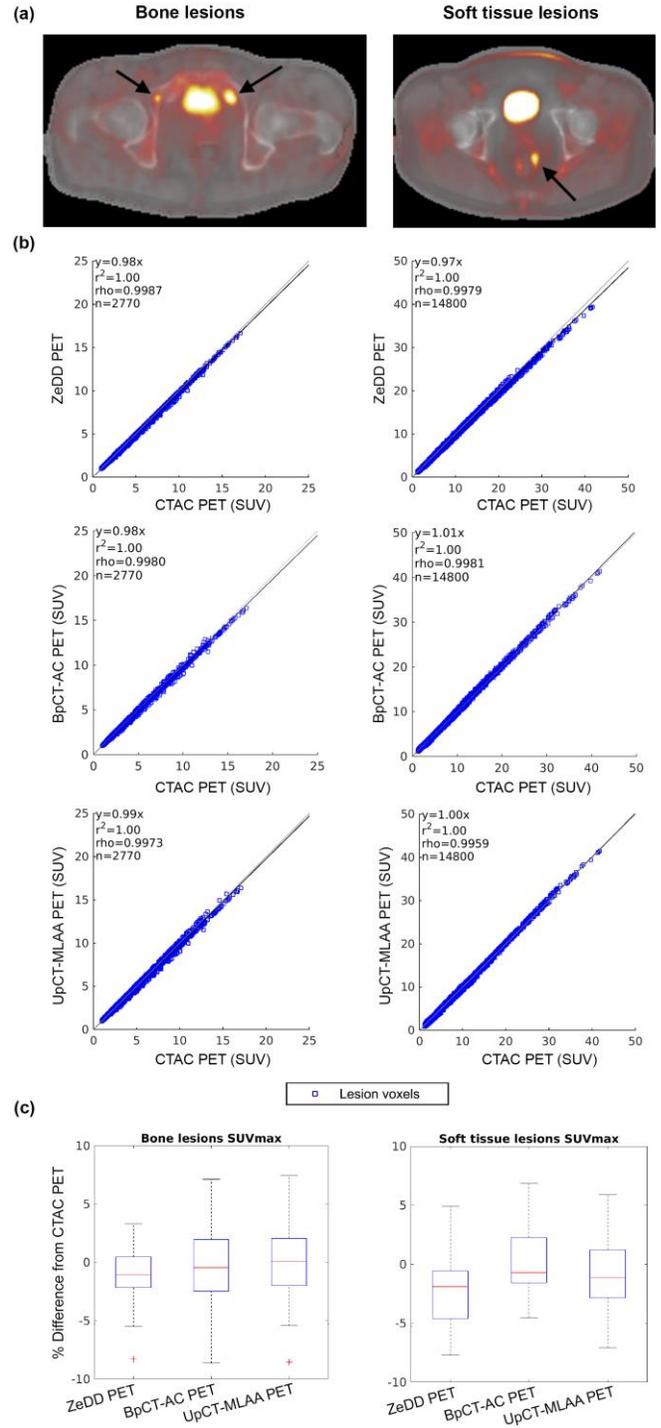

Fig 4. Representative images of bone and soft tissue lesions for patients without implants (A, reproduced from (20)), scatter plots of SUV in every lesion voxel (B), and box plots of the SUVmax in each lesion (C). This shows that BpCT-AC and UpCT-MLAA-AC is near equivalent to ZeDD-CTAC in patients without implants when comparing to CTAC.

### C. PET reconstructions

In addition to UpCT-MLAA, additional PET reconstructions were performed for comparison.

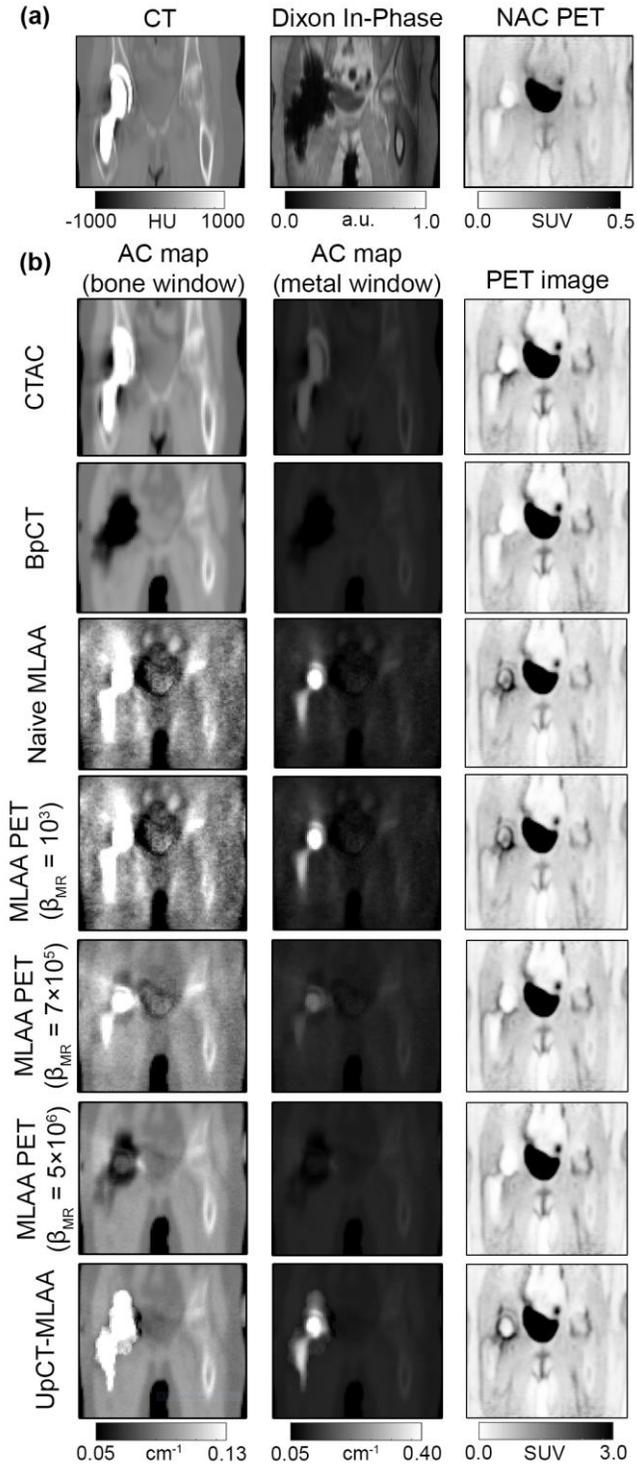

Fig 5. Representative images from metal implant patient #3 imaged with $^{18}$F-FDG. Shown are the CT, Dixon in-phase, and NAC PET images (a), AC maps (b, first and second column), and associated PET reconstructions (b, third column). The AC maps are shown in two different window levels to highlight bone and soft tissue (b, first column) and the metal implant (b, second column).

TABLE I
LESION SUV ERRORS OVER THE VOLUME COMPARED TO CTAC IN PATIENTS WITHOUT IMPLANTS

| *Bone lesions* | |
|---|---|
| Method | SUV RMSE ($\mu \pm \sigma$) |
| ZeDD-CTAC | 2.6 % ($-1.3 \pm 2.3$ %) |
| BpCT-AC | 3.2% ($-0.9 \pm 3.1$%) |
| UpCT-MLAA | 3.6 % ($-0.3 \pm 3.6$ %) |
| Method | SUV$_{max}$ RMSE ($\mu \pm \sigma$) |
| ZeDD-CTAC | 2.6 % ($-1.3 \pm 2.3$ %) |
| BpCT-AC | 3.1% ($-0.3 \pm 3.1$%) |
| UpCT-MLAA | 3.4 % ($0.03 \pm 3.4$ %) |
| *Soft tissue lesions* | |
| Method | SUV RMSE ($\mu \pm \sigma$) |
| ZeDD-CTAC | 4.4 % ($-2.9 \pm 3.3$ %) |
| BpCT-AC | 3.5% ($0.01 \pm 3.5$%) |
| UpCT-MLAA | 4.6 % ($-1.1 \pm 4.5$ %) |
| Method | SUV$_{max}$ RMSE ($\mu \pm \sigma$) |
| ZeDD-CTAC | 4.1 % ($-2.3 \pm 3.4$ %) |
| BpCT-AC | 3.4% ($-0.1 \pm 3.4$%) |
| UpCT-MLAA | 4.8 % ($-1.6 \pm 4.5$ %) |

For each patient without metal implants: (1) UpCT-MLAA was performed and time-of-flight ordered subsets expectation maximization with a point spread function model (TOF-OSEM) [55] (transaxial field of view (FOV) = 600mm, 2 iterations, 28 subsets, matrix size = 192 × 192, 89 slices of 2.78mm thickness) with two $\mu$-maps: (2) ZeDD-CTAC, (3) initial AC estimate of the BCNN (BpCT-AC) and (4) CTAC, for comparison. BpCT-AC is a surrogate for ZeDD-CTAC but without the use of a specialized MR sequence.

For each patient with metal implants, UpCT-MLAA was performed along with (1) naive MLAA, (2 to 4) regularized MLAA with increasing regularization parameters ($\beta_{MR} = [1 \times 10^3, 7 \times 10^5, 5 \times 10^6]$, constant over the volume), (5) TOF-OSEM with BpCT-AC, and (6) TOF-OSEM with CTAC for comparison.

*D. Data Analysis.*

Image error analysis and lesion-based analysis were performed for patients without metal implants: the average ($\mu$) and standard deviation ($\sigma$) of the error, mean-absolute-error (MAE), and root-mean-squared-error (RMSE) were computed over voxels that met a minimum signal amplitude and/or signal-to-noise criteria [21]. Global HU and PET SUV comparisons were only performed in voxels with amplitudes > -950 HU in the ground-truth CT to exclude air, and a similar threshold of > 0.01 cm$^{-1}$ attenuation in the CTAC was used for comparison of AC maps. Bone and soft-tissue lesions were identified by a board-certified radiologist. Bone lesions are defined as lesions inside bone or with lesion boundaries within 10 mm of bone [56]. A Wilcoxon signed-rank test was used to compare the SUV$_{max}$ biases compared to CTAC of individual lesions.

In the cases where a metal implant was present, we qualitatively examined the resulting AC maps of the different reconstructions and quantitatively compared SUV$_{max}$ with reference CTAC PET. High uptake lesions and lesion-like objects were identified on the PET images reconstructed with UpCT-MLAA and separated into two categories: (1) in-plane with the metal implant, and (2) out-plane of the metal implant. A Wilcoxon signed-rank test was used to compare the SUV and SUV$_{max}$ values between the different reconstruction methods and CTAC PET.

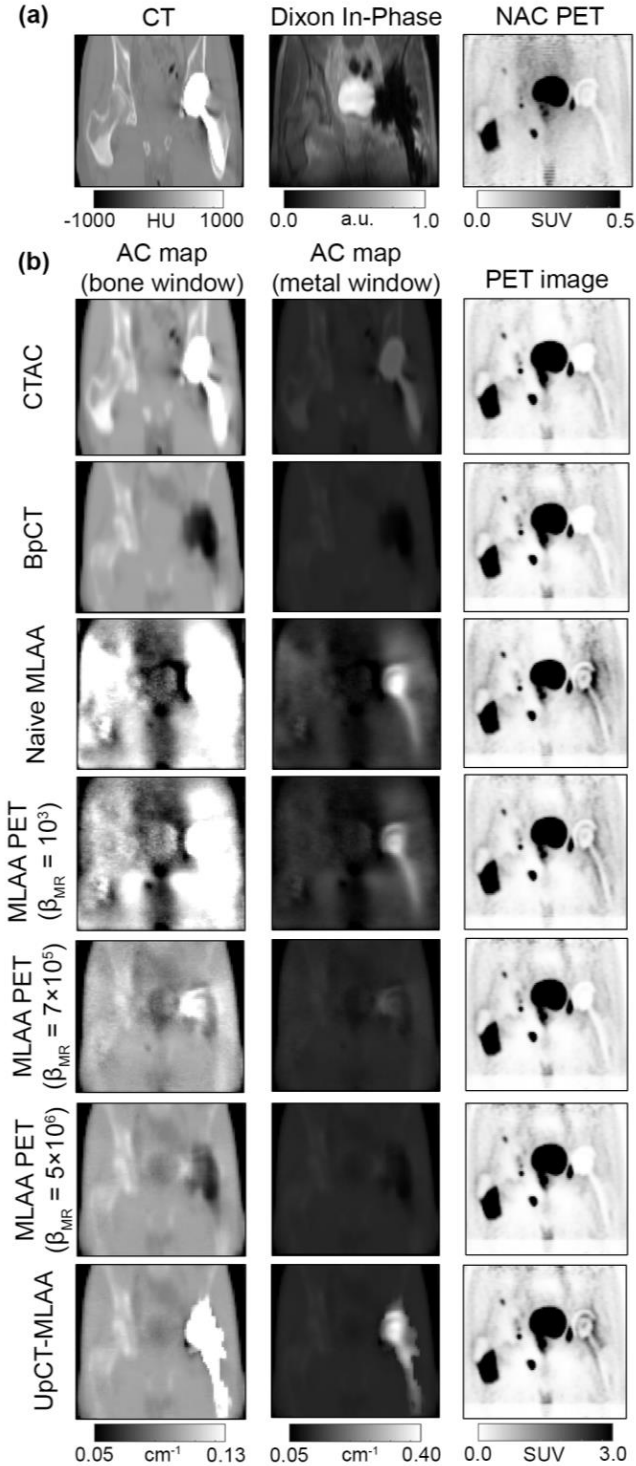

Fig 6. Representative images from metal implant patient #1 imaged with $^{68}$Ga-PSMA. Shown are the CT, Dixon in-phase, and NAC PET images (a), AC maps (b, first and second column), and associated PET reconstructions (b, third column). The AC maps are shown in two different window levels to highlight bone and soft tissue (b, first column) and the metal implant (b, second column).

## IV. RESULTS

### A. Monte Carlo Dropout

Representative images of the output of the BCNN with Monte Carlo Dropout is shown in Fig. 3. The same mask used for the weight maps was used to remove voxels outside the body. The pseudo-CT images visually resemble the ground-truth CT images for patients without implants. While in patients with implants, the metal artifact region in the MRI was assigned air HU values. Nonetheless, the associated standard deviation maps highlighted image structures that the network had high predictive uncertainty. The most important of which are air pockets and the metal implant. The BCNN highlighted these regions and structures in the standard deviation image without being explicitly trained to do so.

An additional example of the uncertainty estimation is provided in Supp. Fig. 1. The input MRI had motion artifacts due to breathing and arm truncation due to inhomogeneity at the edge of the FOV. Like the metal implants, the BCNN highlighted the motion artifact region and arm truncation in the variance image without being explicitly trained to do so.

### B. Patients without implants

The PET reconstruction results for the patients without implants are summarized in Fig. 4. The RMSE is reported along with the average ($\mu$) and standard deviation ($\sigma$) of the error as RMSE ($\mu \pm \sigma$). Additional results for the pseudo-CT, AC maps, and PET data are provided in Supp. Fig. 2 to 5.

*1) Pseudo-CT results*

The total RMSE for the pseudo-CT compared to gold-standard CT across all volumes were 98 HU ($-13 \pm 97$ HU) for ZeDD-CT and 95 HU ($-6.5 \pm 94$ HU) for BpCT. The BpCT is the same pseudo-CT image used in UpCT-MLAA.

*2) Attenuation coefficient map results*

The total RMSE for the AC maps compared to gold-standard CTAC across all volumes were $3.1 \times 10^{-3}$ cm$^{-1}$ ($-5.0 \times 10^{-4} \pm 3.1 \times 10^{-3}$ cm$^{-1}$) for ZeDD-CTAC, $3.2 \times 10^{-3}$ cm$^{-1}$ ($-3.8 \times 10^{-5} \pm 3.2 \times 10^{-3}$ cm$^{-1}$) for BpCT-AC, and $3.5 \times 10^{-3}$ cm$^{-1}$ ($-2.6 \times 10^{-5} \pm 3.5 \times 10^{-3}$ cm$^{-1}$) for UpCT-MLAA-AC.

*3) PET images.*

The total RMSE for PET images compared to gold-standard CTAC PET across all volumes were 0.023 SUV ($-0.005 \pm 0.023$ SUV) for ZeDD PET, 0.022 SUV ($-8.1 \times 10^{-5} \pm 0.022$ SUV) for BpCT-AC PET and 0.027 SUV ($1.5 \times 10^{-4} \pm 0.027$ SUV) for UpCT-MLAA PET.

*4) Lesion uptake and SUV$_{max}$.*

The results for lesion analysis for patients without implants are shown in Fig. 4. There were 30 bone lesions and 60 soft tissue lesions across the 16 patient datasets. The RMSE w.r.t. CTAC PET SUV and SUV$_{max}$ are summarized in Table I. For SUV$_{max}$ of bone lesions, no significant difference was found for ZeDD PET and BpCT-AC PET (p = 0.116) while PET ZeDD PET and UpCT-MLAA PET were significantly different (p = 0.037). For SUV$_{max}$ of soft tissue lesions, ZeDD PET and BpCT-AC PET were significantly different (p < 0.001) while no significant difference was found between ZeDD PET and UpCT-MLAA PET (p = 0.16).

### C. Patients with metal implants

Figs. 5 and 6 show the different AC maps generated with the different reconstruction processes and associated PET image reconstructions on two different radiotracers ($^{18}$F-FDG and $^{68}$Ga-PSMA) and Fig. 7 shows the summary of the SUV$_{max}$ results. Additional results for pseudo-CT, AC maps, and PET images are provided in Supp. Fig. 6 to 11.

*1) Metal implant recovery.*

Figs. 5b (1$^{st}$ and 2$^{nd}$ column) and 6b (1$^{st}$ and 2$^{nd}$ column) show the AC map estimation results.

BpCT-AC filled in the metal implant region with air since the metal artifact in MRI appears as a signal void. Although reconstructing using naive MLAA recovers the metal implant, the AC map was noisy and anatomical structures were difficult to depict. The addition of regularization (increasing $\beta_{MR}$) reduces the noise, however over-regularization eliminates the presence of the metal implant. The use of a different radiotracer also influenced reconstruction performance: the MLAA-based methods performed worse when the tracer was $^{68}$Ga-PSMA compared to $^{18}$F-FDG with low regularization. In contrast, UpCT-MLAA-AC recovered the metal implant while maintaining high SNR depiction of anatomical structures outside the implant region for both radiotracers. The high attenuation coefficients were constrained in the regions where high variance was measured (or where the metal artifact was present on the BpCT AC maps).

*2) PET image reconstruction*

Figs. 5b (3$^{rd}$ column) and 6b (3$^{rd}$ column) show the PET image reconstructions results.

Qualitatively, the MLAA-based methods (UpCT-MLAA and Standard MLAA) show uptake around the implant, whereas BpCT-AC PET and CTAC PET show the implant region without any uptake. When compared to the NAC PET, the MLAA-based methods better match what is depicted within the implant region. Quantitatively, Table I summarizes the SUV results for voxels in-plane of the metal implant and out-plane of the metal implant.

*3) SUV$_{max}$ quantification.*

Fig. 7 shows the comparisons of SUV$_{max}$ of lesions in-plane and out-plane of the metal implant and Table II and Table III list the RMSE values for SUV and SUV$_{max}$. There were 6 lesions in-plane and 15 lesions out-plane with the metal implants across the 3 patients with implants. Only UpCT-MLAA provided relatively low SUV$_{max}$ quantification errors on lesions both in-plane and out-plane of the metal implant.

For lesions in-plane of the metal implant, BpCT-AC PET had large underestimation of SUV$_{max}$, naive MLAA PET had better mean estimation of SUV$_{max}$ but had a large standard deviation. The addition of light regularization to MLAA improves the RMSE by decreasing the standard deviation at the cost of increased mean error. Increasing regularization increases RMSE but reduces the bias error with increased standard deviation. UpCT-MLAA PET had the best agreement with CTAC PET. Only Naive MLAA and UpCT-MLAA had results where a significant difference could not be found when compared to CTAC (p > 0.05).

For lesions out-plane of the metal implant, the trend is reverse for BpCT-AC PET and the MLAA methods. BpCT-AC PET had the best agreement with CTAC PET and the MLAA methods showed decreasing RMSE with increasing regularization. UpCT-MLAA had the second-best agreement with CTAC PET. No significant difference could be found for all methods when compared to CTAC (p > 0.05)

## V. DISCUSSION

This paper presents the use of a Bayesian deep convolutional neural network to enhance MLAA by providing an accurate pseudo-CT prior alongside predictive uncertainty estimates that automatically modulate the strength of the priors (UpCT-MLAA). The method was evaluated in patients without and with implants with pelvic lesions. The performance for metal implant recovery and uptake estimation in pelvic lesions in patients with metal implants was characterized. This is the first work that demonstrated an MLAA algorithm for PET/MRI that was able to recover metal implants while also accurately depicting detailed anatomic structures in the pelvis. This is also the first work to synergistically combine supervised Bayesian deep learning and MLAA in a coherent framework for simultaneous PET/MRI reconstruction in the pelvis. The UpCT-MLAA method demonstrated similar quantitative uptake estimation of pelvic lesions to a state-of-the-art attenuation correction method (ZeDD-CT) while additionally providing the capability to perform reasonable PET reconstruction in the presence of metal implants and removing the need of a specialized MR pulse sequence.

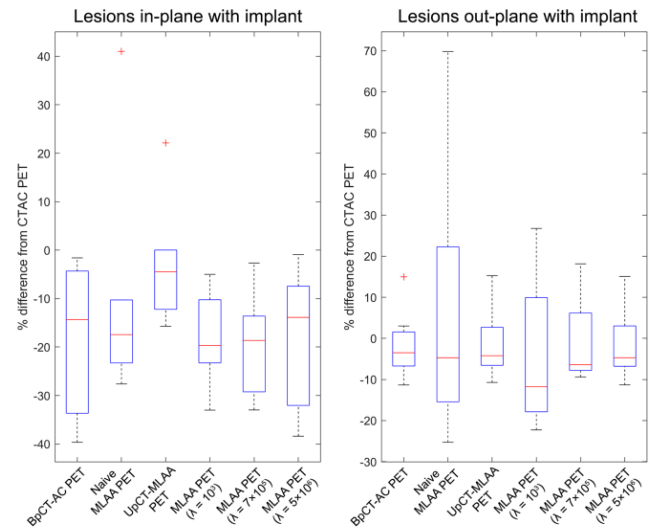

Fig 7. Box plot summarizing the results comparing to CTAC PET for patients with implants. The red crosses denote outliers.

TABLE II
SUV ERRORS OVER THE VOLUME COMPARED TO CTAC

| Method | SUV RMSE ($\mu, \sigma$, p-value) |
|---|---|
| *In-plane with metal implant* | |
| BpCT-AC | 0.10 SUV ($\mu = -0.014$ SUV, $\sigma = 0.10$ SUV, $p < 0.001$) |
| Naive MLAA | 0.19 SUV ($\mu = 0.06$ SUV, $\sigma = 0.18$ SUV, $p < 0.001$) |
| MLAA PET ($\beta_{MR} = 10^3$) | 0.16 SUV ($\mu = 0.047$, $\sigma = 0.15$, $p < 0.001$) |
| MLAA PET ($\beta_{MR} = 7 \times 10^5$) | 0.09 SUV ($\mu = 0.001$, $\sigma = 0.09$, $p < 0.001$) |
| MLAA PET ($\beta_{MR} = 5 \times 10^6$) | 0.09 SUV ($\mu = -0.008$, $\sigma = 0.09$, $p < 0.001$) |
| UpCT-MLAA | 0.12 SUV ($\mu = 0.018$ SUV, $\sigma = 0.12$ SUV, $p < 0.001$) |
| *Out-plane with metal implant* | |
| BpCT-AC | 0.086 SUV ($\mu = 0.009$ SUV, $\sigma = 0.085$ SUV, $p < 0.001$) |
| Naive MLAA | 0.14 SUV ($\mu = 0.018$ SUV, $\sigma = 0.14$ SUV, $p < 0.001$) |
| MLAA PET ($\beta_{MR} = 10^3$) | 0.13 SUV ($\mu = 0.016$, $\sigma = 0.13$, $p < 0.001$) |
| MLAA PET ($\beta_{MR} = 7 \times 10^5$) | 0.09 SUV ($\mu = 0.012$, $\sigma = 0.09$, $p < 0.001$) |
| MLAA PET ($\beta_{MR} = 5 \times 10^6$) | 0.09 SUV ($\mu = 0.010$, $\sigma = 0.09$, $p < 0.001$) |
| UpCT-MLAA | 0.086 SUV ($\mu = 0.012$ SUV, $\sigma = 0.085$ SUV, $p < 0.001$) |

TABLE III
LESION SUV$_{MAX}$ PERCENT ERRORS

| Method | SUV$_{max}$ % RMSE ($\mu, \sigma$, p-value) |
|---|---|
| *In-plane with metal implant (n=6)* | |
| BpCT-AC | 24.2% ($\mu = -18.0\%$, $\sigma = 16.2\%$, $p = 0.03$) |
| Naive MLAA | 26.9% ($\mu = -9.2\%$, $\sigma = 25.3\%$, $p = 0.31$) |
| MLAA PET ($\beta_{MR} = 10^3$) | 21.0% ($\mu = -18.5\%$, $\sigma = 9.9\%$, $p = 0.03$) |
| MLAA PET ($\beta_{MR} = 7 \times 10^5$) | 22.2% ($\mu = -19.3\%$, $\sigma = 10.9\%$, $p = 0.03$) |
| MLAA PET ($\beta_{MR} = 5 \times 10^6$) | 23.1% ($\mu = -17.8\%$, $\sigma = 14.8\%$, $p = 0.03$) |
| UpCT-MLAA | 13.6% ($\mu = -2.4\%$, $\sigma = 13.4\%$, $p = 0.44$) |
| *Out-plane with metal implant (n=15)* | |
| BpCT-AC | 6.9% ($\mu = -2.5\%$, $\sigma = -6.5\%$, $p = 0.07$) |
| Naive MLAA | 27.9% ($\mu = 4.7\%$, $\sigma = 27.5\%$, $p = 0.72$) |
| MLAA PET ($\beta_{MR} = 10^3$) | 18.7% ($\mu = -5.2\%$, $\sigma = 18.0\%$, $p = 0.28$) |
| MLAA PET ($\beta_{MR} = 7 \times 10^5$) | 9.6% ($\mu = -1.3\%$, $\sigma = 9.5\%$, $p = 0.33$) |
| MLAA PET ($\beta_{MR} = 5 \times 10^6$) | 7.4% ($\mu = -2.1\%$, $\sigma = 7.1\%$, $p = 0.21$) |
| UpCT-MLAA | 7.1% ($\mu = -1.9\%$, $\sigma = 6.8\%$, $p = 0.19$) |

One of the major advantages of using MLAA is that it uses the PET emission data to estimate the attenuation coefficients alongside the emission activity. This gives MLAA the capability to truly capture the underlying imaging conditions that the PET photons undergo. This is especially important in simultaneous PET/MRI where true ground-truth attenuation maps cannot be derived. Currently, the most successful methods for obtaining attenuation maps are through deep learning-based methods [20]–[28]. However, these methods are inherently supervised model-based techniques and have limited capacity to capture imaging conditions that were not present in the training set nor conditions that cannot be reliably modeled, such as the movement and mismatch of bowel air and the presence of metal artifacts. Since MLAA derives the attenuation maps from the PET emission data, MLAA can derive actual imaging conditions that supervised model-based techniques are unable to capture. Furthermore, this eliminates the need for specialized MR pulse sequence (such as ZTE for bone) since the bone AC would be estimated by MLAA instead. This would allow for more accurate and precise uptake quantification in simultaneous PET/MRI.

To the best of our knowledge, only a few other methods combines MLAA with deep learning [39]–[42]. Their methods apply deep learning to denoise an MLAA reconstruction by training a deep convolutional neural network to produce an equivalent CTAC from MLAA estimates of activity and attenuation maps. This method inherently requires ground-truth CTAC maps to train the deep convolutional neural network and thus is affected by the same limitations that supervised deep learning and model-based methods have. Unlike their method, our method (UpCT-MLAA) preserves the underlying MLAA reconstruction while still providing the same reduction of crosstalk artifacts and noise.

Our approach is different from all other approaches because we leverage supervised Bayesian deep learning uncertainty estimation to detect rare and previously unseen structures in pseudo-CT estimation. There are only a few previous works that estimate uncertainty on pseudo-CT generation [57], [58]. Klages et al [57] utilized a standard deep learning approach and extracted patch uncertainty but did not assess their method on cases with artifacts or implants. Hemsley et al [58] utilized a Bayesian deep learning approach to estimate total predictive uncertainty and similarly demonstrated high uncertainty on metal artifacts. Both approaches were intended for radiotherapy planning and our work is the first to apply uncertainty estimation towards PET/MRI attenuation correction. We demonstrated how likely $\mu$-map errors can be detected and resolved with the use of PET emission data through MLAA.

High uncertainty was present in many different regions. Metal artifact regions had high uncertainty because they were explicitly excluded in the training process—i.e., an out-of-distribution structure. Air pockets had high uncertainty likely because of the inconsistent correspondence of air between MRI and CT—i.e., intrinsic dataset errors. Other image artifacts (such as motion due to breathing) have high uncertainty likely due to the rare occurrence of these features in the training dataset and its inconsistency with the corresponding CT images. Bone had high uncertainty since there is practically no bone signal in the Dixon MRI. Thus, the CNN likely learned to derive bone value based on the surrounding structure and the variance image shows the intrinsic uncertainty and limitations of estimating bone HU values from Dixon MRI. Again, these regions were highlighted by being assigned high uncertainty without the network being explicitly trained to identify these regions.

On evaluation with patients without implants, we demonstrated that BpCT was a sufficient surrogate of ZeDD-CT for attenuation correction across all lesion types: BpCT provided comparable SUV estimation on bone lesions and improved SUV estimation on soft tissue lesions. However, the BpCT images lacked accurate estimation of bone HU values that resulted in average underestimation of bone lesion SUV values (-0.9%). The average underestimation was reduced with UpCT-MLAA (-0.3%). Although the mean underestimation values improved, the RMSE of UpCT-MLAA was higher than BpCT-AC (3.6% vs. 3.2%, respectively) due to the increase in standard deviation (3.6% vs. 3.1%, respectively). This trend was more apparent for soft tissue lesions. The RMSE, mean error, and standard deviation were worse for UpCT-MLAA vs. BpCT. Since the PET/MRI and CT were acquired in separate sessions, possibly months apart, there may be significant changes in tissue distribution. This could explain the increase in errors of BpCT-AC under UpCT-MLAA.

On the patients with metal implants, UpCT-MLAA was the most comparable to CTAC across all lesion types. Notably, there was an opposing trend in the PET $SUV_{max}$ results for lesions in/out-plane of the metal implant between BpCT-AC and the MLAA methods. These were likely due to the sources of data for reconstruction. BpCT-AC has attenuation coefficients estimated only from the MRI whereas Naïve MLAA has attenuation coefficients estimated only from the PET emission data. The input MRI were affected by large metal artifacts due to the metal implants that makes the regions appear to be large pockets of air. Thus, in BpCT-AC, the attenuation coefficients of air were assigned to the metal artifact region. For lesions in-plane of the implant, this led to a large bias due to the bulk error in attenuation coefficients and a large variance due to the large range of attenuation coefficients with BpCT-AC, while this is resolved with MLAA. For lesions out-plane of the implant, the opposite trend arises. For MLAA the variance is large due to the noise in the attenuation coefficient estimates. This is resolved in BpCT-AC since the attenuation coefficients are learned for normal anatomical structures that are unaffected by metal artifacts. The combination of BpCT with MLAA through UpCT-MLAA resolved these disparities.

A major challenge to evaluate PET reconstructions in the presence of metal implants is that typical CT protocols for CTAC produce metal implant artifacts that may cause overestimation of uptake and thus does not serve as a true reference. Since our method relies on time-of-flight MLAA, we believe that our method would produce a more accurate AC map, and therefore more accurate SUV map. This is demonstrated by the lower $SUV_{max}$ estimates of UpCT-MLAA compared to CTAC PET. However, to have precise evaluation, a potential approach to evaluate UpCT-MLAA is to use metal artifact reduction techniques on the CT acquisition [43] or by acquiring transmission PET images [59].

Accurate co-registration of CT and MRI with metal implant artifacts was a limitation since the artifacts present themselves differently. Furthermore, the CT and MRI images were acquired in separate sessions. These can be mitigated by acquiring images sequentially in a tri-modality system [60].

Another limitation of this study was the small study population. Having a larger population would allow evaluation with a larger variety of implant configurations and radiotracers and validation of the robustness of the attenuation correction strategy.

Finally, the performance of the algorithm can be further improved. In this study, we only sought to demonstrate the utility of uncertainty estimation with a Bayesian deep learning regime for the attenuation correction in the presence of metal implants: that the structure of the anatomy is preserved and implants can be recovered while still providing similar PET uptake estimation performance in pelvic lesions. Our proposed UpCT-MLAA was based on MLAA regularized with MR-based priors [27], which can be viewed as uni-modal Gaussian priors. We speculate that this could be further improved by using Gaussian mixture priors for MLAA as in [36]. The major task to combine these methods would be to learn the Gaussian mixture model parameters from patients with implants. With additional tuning of the algorithm and optimization of the BCNN, UpCT-MLAA can potentially produce the most accurate and precise attenuation coefficients in all tissues and in any imaging conditions.

## VI. CONCLUSION

We have developed and evaluated an algorithm that utilizes a Bayesian deep convolutional neural network that provides accurate pseudo-CT priors with uncertainty estimation to enhance MLAA PET reconstruction. The uncertainty estimation allows for the detection of "out-of-distribution" pseudo-CT estimates that MLAA can subsequently correct. We demonstrated quantitative accuracy in pelvic lesions and recovery of metal implants in pelvis PET/MRI.

**Supplementary Material for "Attenuation Coefficient Estimation for PET/MRI With Bayesian Deep Learning pseudo-CT and Maximum Likelihood Estimation of Activity and Attenuation" by A. P. Leynes et al.**

*Sources of uncertainty and variations*

There are three different predictive uncertainties that are utilized in our work: total voxel uncertainty--and its components--patch uncertainty, and voxel-wise uncertainty.

Total voxel uncertainty is the combination of patch uncertainty (uncertainty due to changes in input patch) and uncertainty of each voxel for the same input patch (uncertainty due to changes in the model). These can be decoupled and independently estimated.

Patch uncertainty comes from variations of the response of the CNN due to changes in the input data. Whereas voxel uncertainty (for the same input patch) come from variations of the network parameters with respect to the same input. Mathematically, the predictive likelihood for a single voxel can be written completely as follows:

$$p(y_i^* | \mathbf{X}^*, \mathbf{X}, \mathbf{Y}) = \frac{1}{N} \sum_{x^* \in \mathbf{X}^*} \int p(y^*|x^*, \theta) p(\theta|\mathbf{X}, \mathbf{Y}) d\theta$$

where $y_i^*$ is the predicted value at the $i$-th voxel, $\mathbf{X}^*$ is the set of neighboring and overlapping input patches, N is the number of patches to predict the value of a single voxel, $\theta$ are the network parameters, $\mathbf{X}, \mathbf{Y}$ are the training input/output pairs, and $p(\theta|\mathbf{X}, \mathbf{Y})$ is the posterior distribution of the network parameters given the training pairs that is learned during model training.

The final predicted value is obtained by taking the expectation of the model predictions over the predictive likelihood:

$$\widehat{y_i^*} = E[y_i^*] = \int y_i^* \left( \frac{1}{N} \sum_{x^* \in \mathbf{X}^*} \int p(y^*|x^*, \theta) p(\theta|\mathbf{X}, \mathbf{Y}) d\theta \right) dy_i^*$$

$$\approx \frac{1}{NM} \sum_{x^* \in \mathbf{X}^*} \sum_{m=1}^{M} y_{i,m}^* \quad (By\ Monte-Carlo\ approximation\ of\ the\ integrals)$$

$$= \frac{1}{NM} \sum_{x^* \in \mathbf{X}^*} \sum_{m=1}^{M} f_{CNN}(x^*, \theta_m)$$

And the variance is:

$$\sigma_{y_i^*}^2 = E[(y_i^* - \widehat{y_i^*})^2] = \int (y_i^* - \widehat{y_i^*})^2 \left( \frac{1}{N} \sum_{x^* \in \mathbf{X}^*} \int p(y^*|x^*, \theta) p(\theta|\mathbf{X}, \mathbf{Y}) d\theta \right) dy_i^*$$

$$\approx \frac{1}{NM} \sum_{x^* \in \mathbf{X}^*} \sum_{m=1}^{M} (y_{i,m}^* - \widehat{y_i^*})^2 \quad (By\ Monte-Carlo\ approximation\ of\ the\ integrals)$$

$$= \frac{1}{NM} \sum_{x^* \in \mathbf{X}^*} \sum_{m=1}^{M} (f_{CNN}(x^*, \theta_m) - \widehat{y_i^*})^2$$

where M is the number of Monte-Carlo samples used in inference.

Voxel uncertainty corresponds to the following term in the predictive likelihood:

$$\int p(y^*|x^*, \theta) p(\theta|X, Y) d\theta$$

and corresponds to the following summation in the prediction and variance:

$$\sum_{m=1}^{M} f_{CNN}(x^*, \theta_m)$$

$$\sum_{m=1}^{M} (f_{CNN}(x^*, \theta_m) - \widehat{y_i^*})^2$$

The patch uncertainty comes from averaging the predictions of different input patches for each single voxel and corresponds to the summation in the prediction and variance:

$$\sum_{x^* \in X^*} f_{CNN}(x^*, \theta_m)$$

$$\sum_{x^* \in X^*} (f_{CNN}(x^*, \theta_m) - \widehat{y_i^*})^2$$

Suppose that there is no model uncertainty, and the network parameters are fixed to be $\hat{\theta}$ (only one set of network parameters used in all inferences). The predictive likelihood will then be

$$p(y_i^*|X^*, X, Y) = \frac{1}{N} \sum_{x^* \in X^*} p(y^*|x^*, \hat{\theta})$$

And the final predicted value will be:

$$\widehat{y_i^*} = \frac{1}{N} \sum_{x^* \in X^*} f_{CNN}(x^*, \hat{\theta})$$

And the variance will be

$$\sigma_{y_i^*}^2 = \frac{1}{N} \sum_{x^* \in X^*} (f_{CNN}(x^*, \hat{\theta}) - \widehat{y_i^*})^2$$

Suppose that we do not process overlapping patches and only extract voxel uncertainty, the predictive likelihood will be:

$$p(y_i^*|x^*, X, Y) = \int p(y^*|x^*, \theta) p(\theta|X, Y) d\theta$$

And the final predicted value will be:

$$\widehat{y_i^*} = \frac{1}{M} \sum_{m=1}^{M} f_{CNN}(x^*, \theta_m)$$

And the variance will be:

$$\sigma^2_{y^*_i} = \frac{1}{M} \sum_{m=1}^{M} (f_{CNN}(x^*, \theta_m) - \widehat{y^*_i})^2$$

Thus, patch uncertainty and (patch-specific) voxel uncertainty can be independently obtained but are tightly coupled together when calculating total voxel uncertainty. In the final prediction for this work, we utilize total voxel uncertainty that incorporates both patch uncertainty and (patch-specific) voxel uncertainty.

*Additional information regarding the weight map empirical transformation*

The empirical transformation was a heuristic developed for this work to be compatible with the MLAA algorithm of Ahn et al [1]. In the work of Ahn et al, the weight prior, $\beta_{MR}$, was defined only by two discrete values: one for the implant region and one for outside the implant region. Our work maintained the use of upper and lower bounds and we used a smooth curve that allows for smoothness as the input values goes towards the saturation values. The predicted variance values have a range of $[0, \infty)$. We set an upper-bound threshold for variance values based on the observed ranges and visual inspection of which anatomic structures the high-variance regions corresponded to. The complete transformation function then has the following requirement: $f: [0, \sigma^2_{max}] \rightarrow [\beta_{MR_{min}}, \beta_{MR_{max}}]$ and we desired a smooth transition towards the saturation values. Thus, we chose a sigmoid function for initial transformation the transformation of the variance values: $f: [0, \sigma^2_{max}] \rightarrow [0, 1]$, and the linear transformation re-maps it to the range needed for $\beta_{MR}$: $g: [0, 1] \rightarrow [\beta_{MR_{min}}, \beta_{MR_{max}}]$.

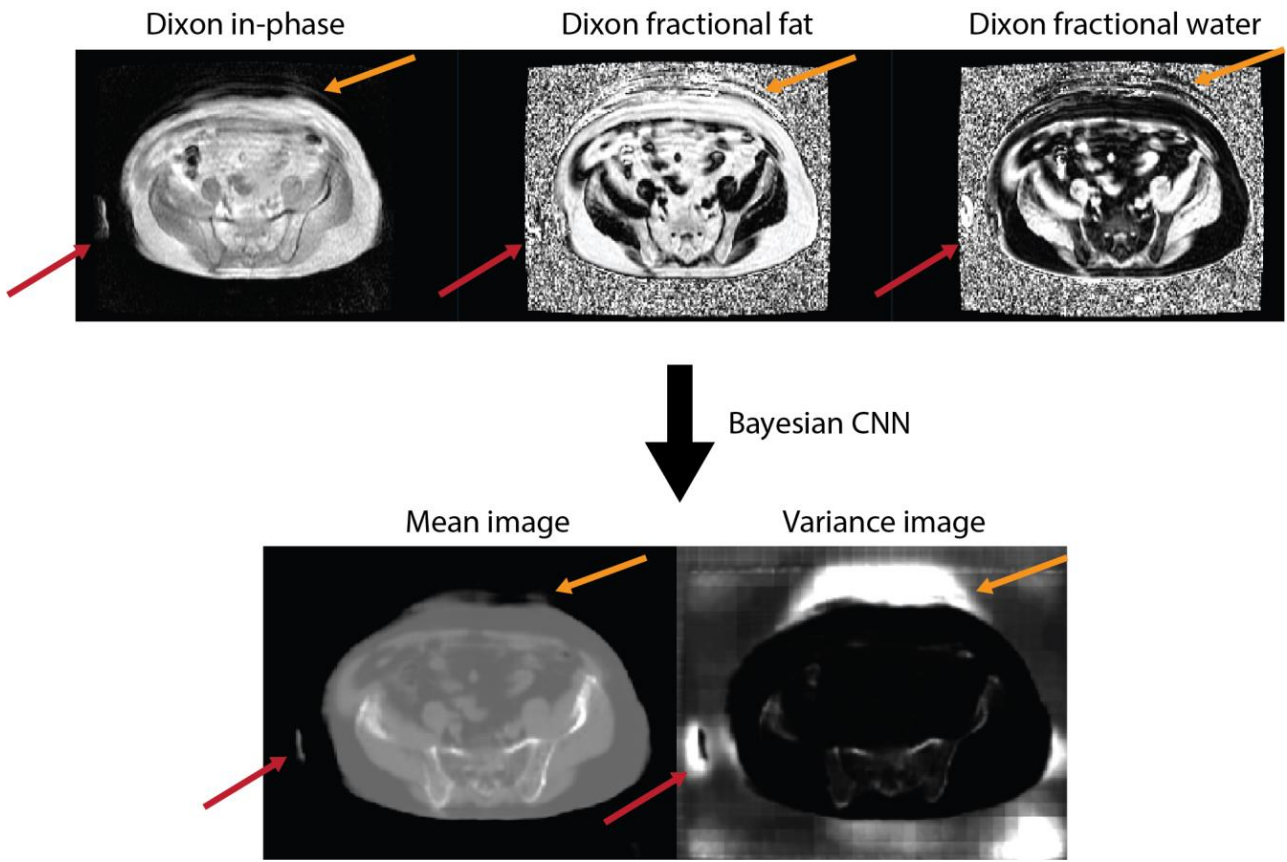

Supp. Fig. 1. Result of uncertainty estimation on without body masks on an MRI with motion artifacts due to breathing and arm truncation due to edge of field-of-view inhomogeneity. The orange arrow points to a region where motion artifacts were present and the dark red arrow points to a region with arm truncation. Both artifact regions were highlighted in the variance image without the network being explicitly trained to highlight these regions.

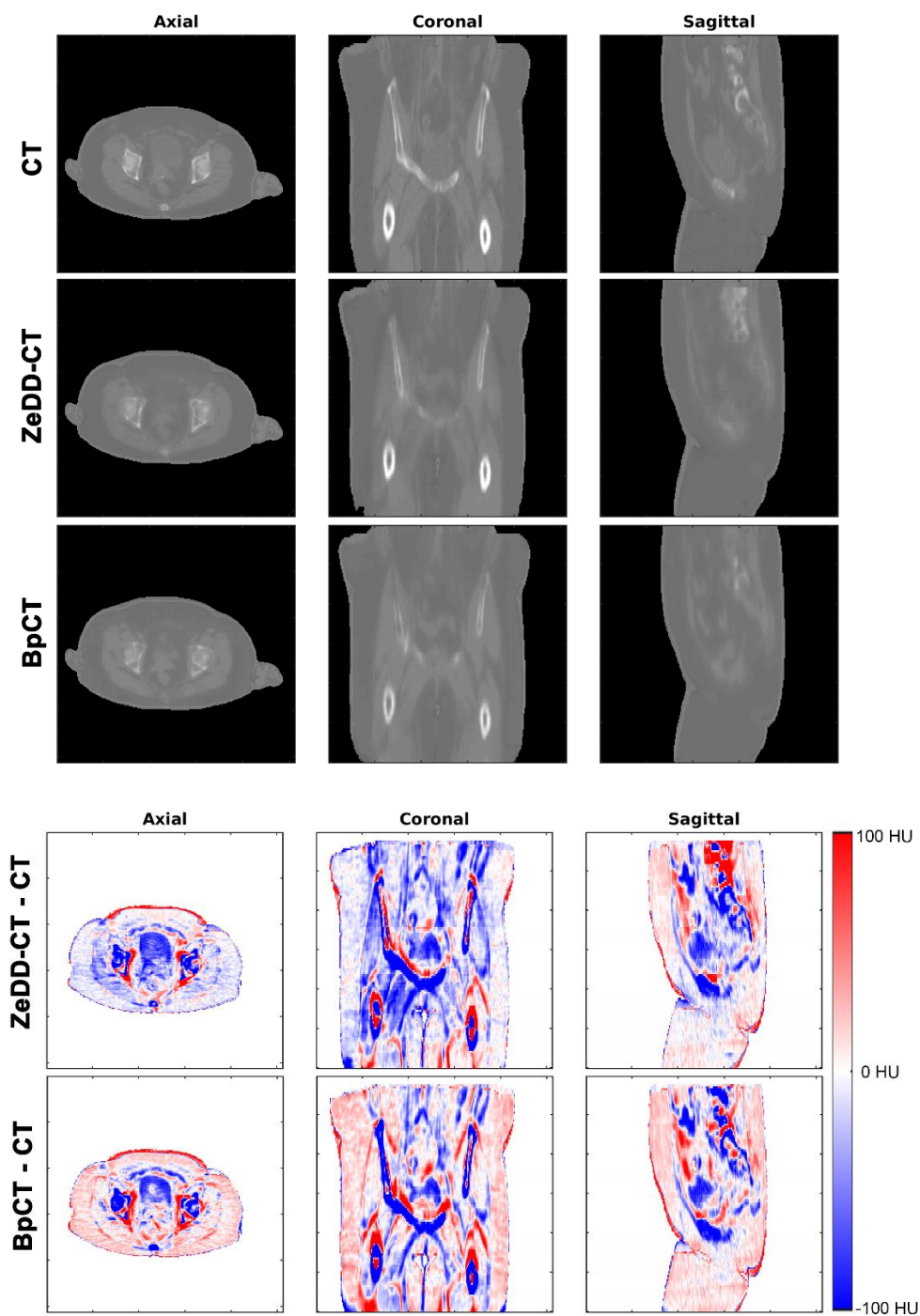

Supp. Fig. 2. CT and difference images of pseudo-CT images for one representative case without implants. Both ZeDD-CT and BpCT have the least RMSE for this patient (RMSE = 98.5 HU, and RMSE = 98.9 HU, respectively).

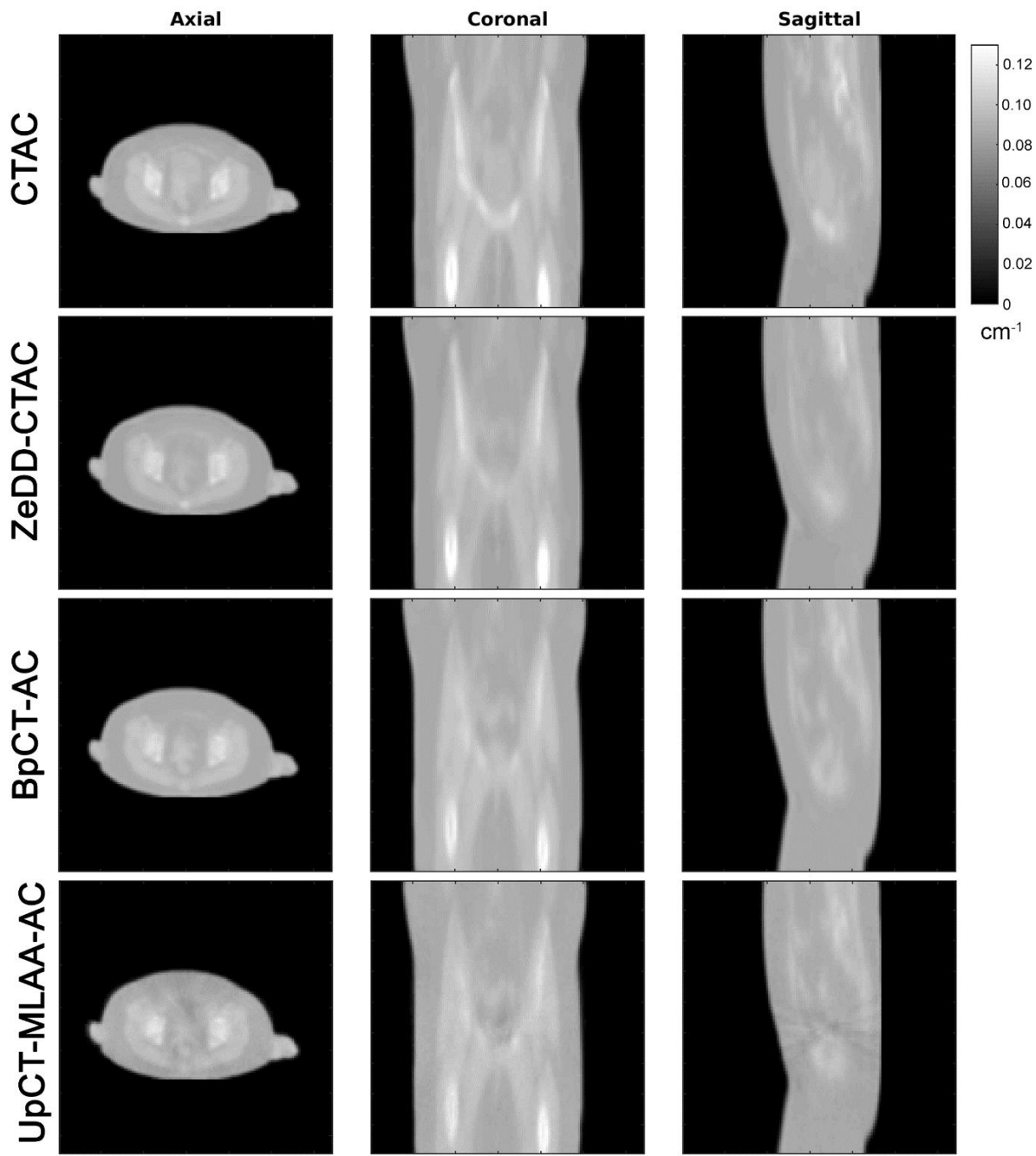

Supp. Fig. 3. CTAC and the different AC maps produced from the different methods for one representative case without implants.

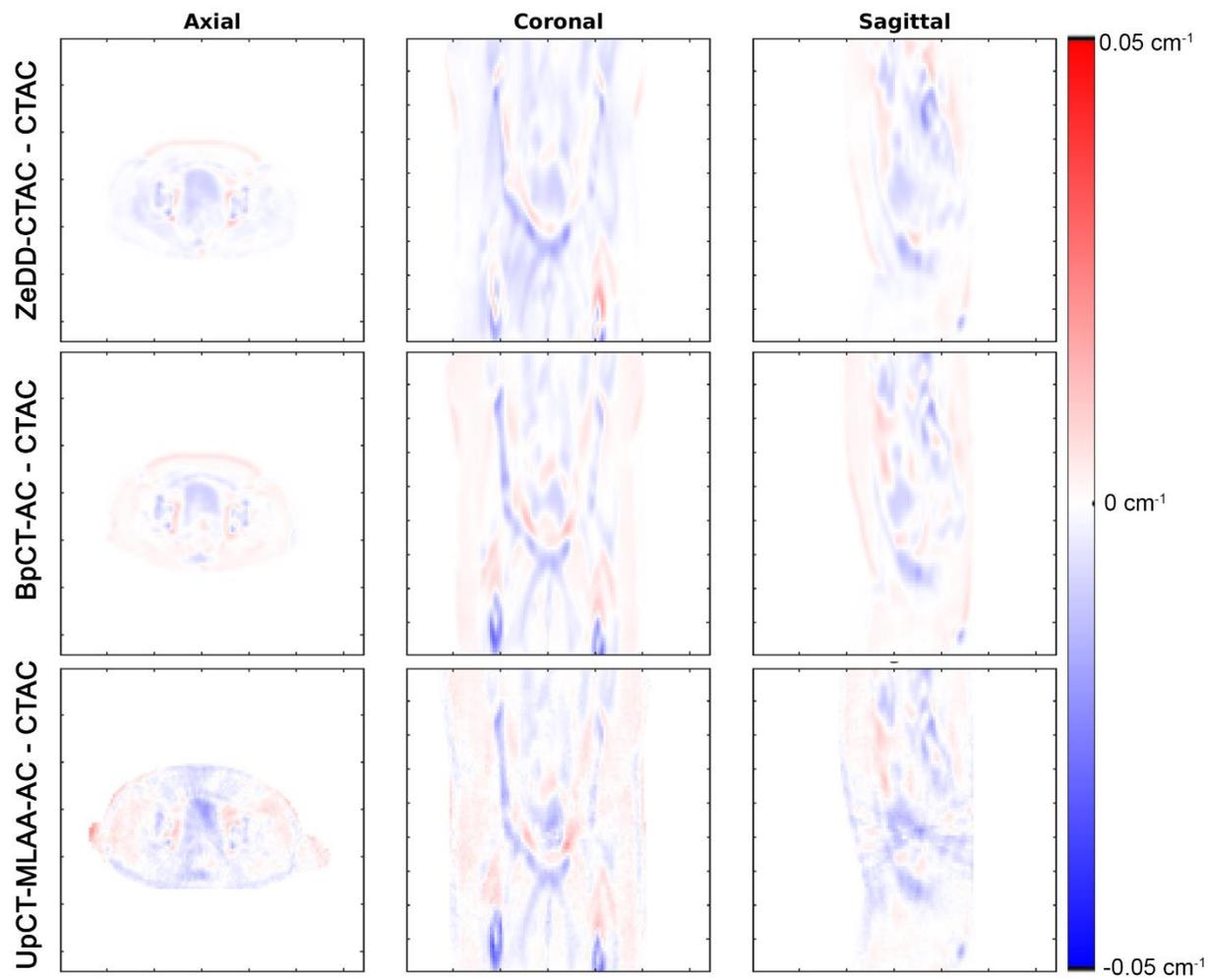

Supp. Fig. 4. Difference images of the AC methods compared to ground-truth CTAC for one representative case without implants.

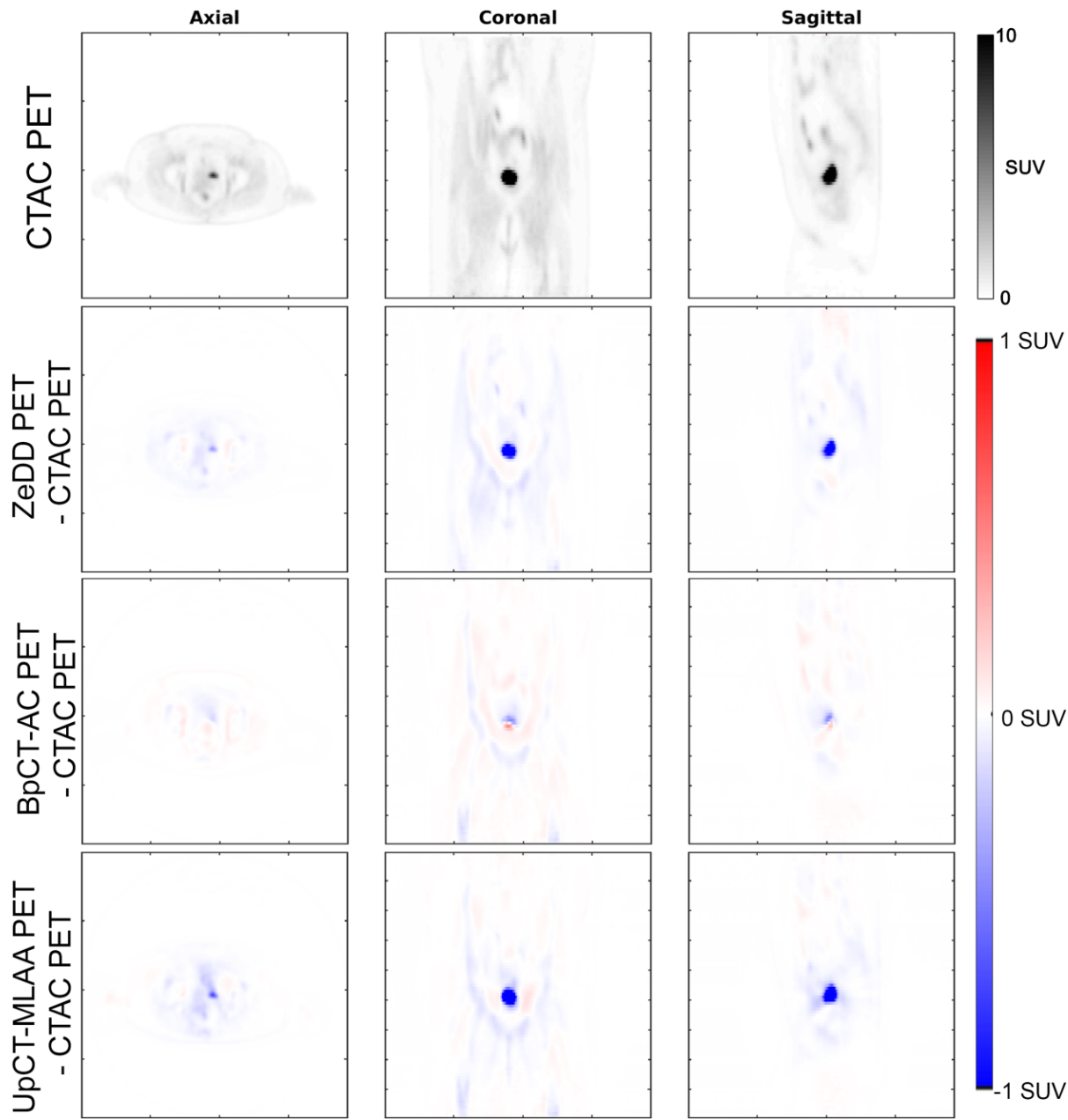

Supp. Fig. 5. PET images of CTAC PET and difference images of the AC methods compared to CTAC PET for one representative case without implants.

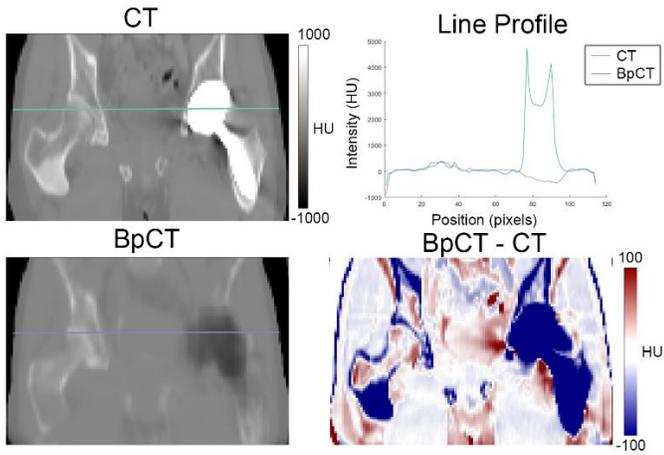

Supp. Fig. 6. CT, pseudo-CT, line profiles, and difference images for a patient with a metal implant imaged with PSMA.

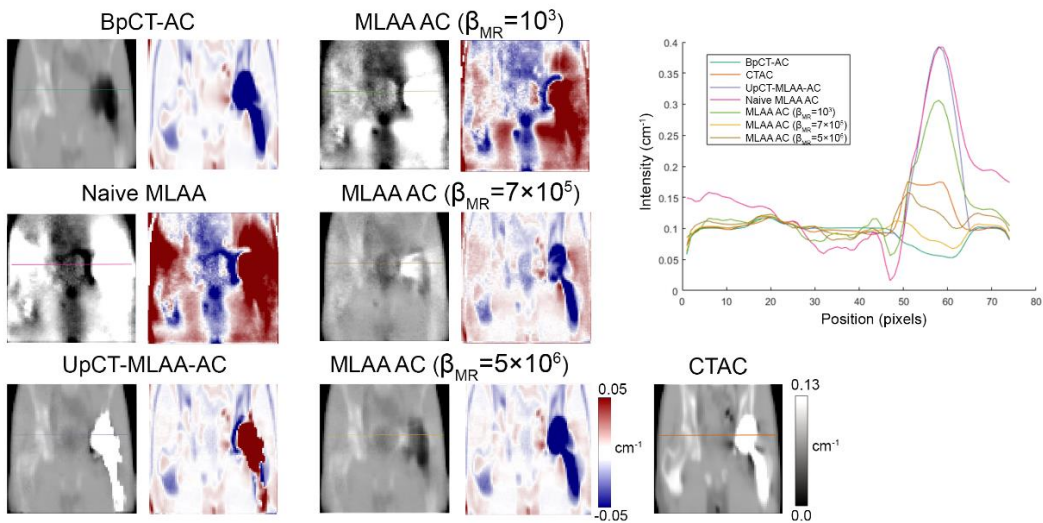

Supp. Fig. 7. AC maps, line profile, and difference images for a patient with a metal implant imaged with PSMA.

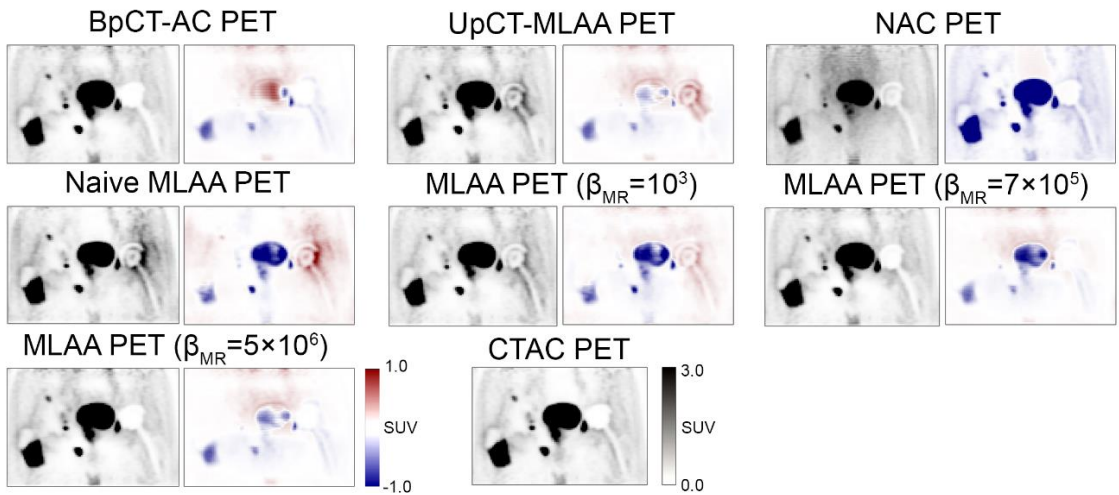

Supp. Fig. 8. PET images and difference images for a patient with a metal implant imaged with PSMA.

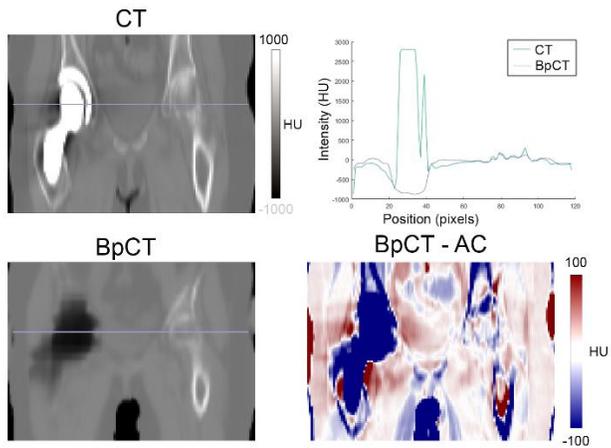

Supp. Fig. 9. CT, pseudo-CT, line profiles, and difference images for a patient with a metal implant imaged with FDG.

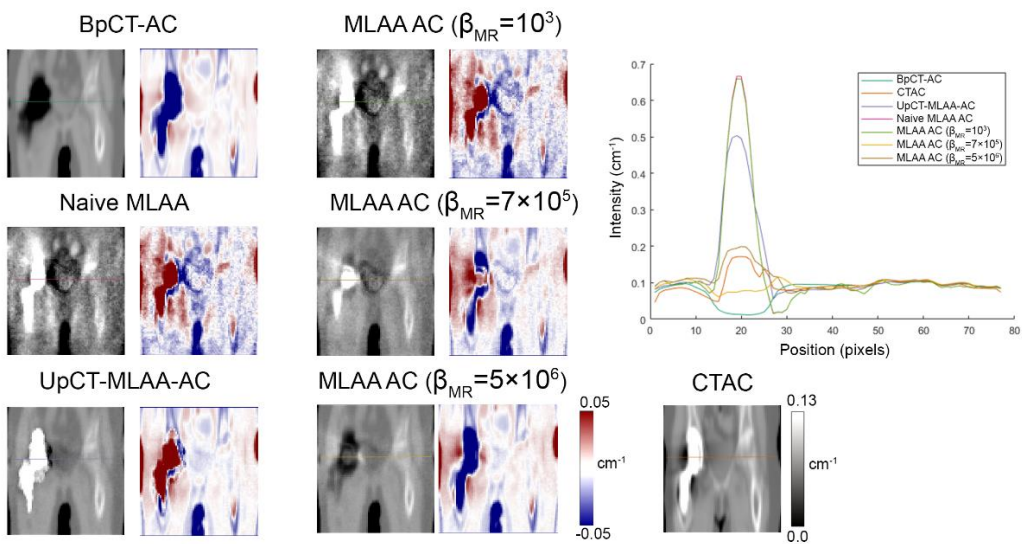

Supp. Fig. 10. AC maps, line profile, and difference images for a patient with a metal implant imaged with FDG.

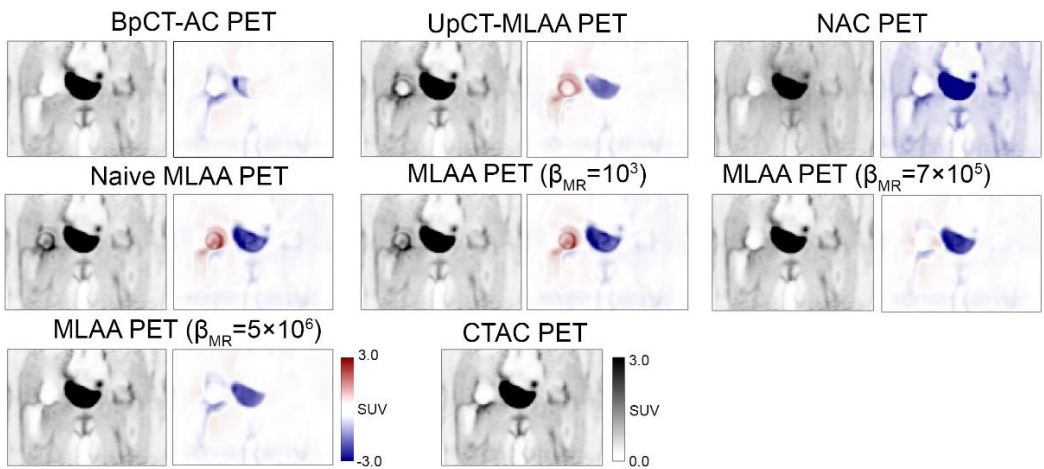

Supp. Fig. 11. PET images and difference images for a patient with a metal implant imaged with FDG.